\documentclass[pre,aps,twocolumn,superscriptaddress,floatfix]{revtex4-1}

\usepackage{amsmath}
\usepackage{amssymb}
\usepackage{amsbsy}
\usepackage{graphicx}
\usepackage{color}
\usepackage{bigints} 
\usepackage{enumerate}
\usepackage{mathtools}
\usepackage{ulem}

\def\be{\begin{equation}}
\def\ee{\end{equation}}
\def\bfi{\begin{figure}}      
\def\efi{\end{figure}}
\def\bea{\begin{eqnarray}}
\def\eea{\end{eqnarray}}

\begin{document}

\title{Getting hotter by heating less: \\ How driven granular materials dissipate energy in excess}

\author{A. Plati}
\affiliation{Department of Physics, University of Rome Sapienza, P.le Aldo Moro 2, 00185, Rome, Italy}
\affiliation{Institute for Complex Systems - CNR, P.le Aldo Moro 2, 00185, Rome, Italy}

\author{L. de Arcangelis}
\affiliation{Department of Engineering, University of Campania ``Luigi Vanvitelli'',  81031, Aversa (Caserta), Italy}

\author{A. Gnoli}
\affiliation{Institute for Complex Systems - CNR, P.le Aldo Moro 2, 00185, Rome, Italy}
\affiliation{Department of Physics, University of Rome Sapienza, P.le Aldo Moro 2, 00185, Rome, Italy}

\author{E. Lippiello}
\affiliation{Department of Mathematics and Physics, University of Campania ``Luigi Vanvitelli'', 81100, Caserta, Italy}

\author{A. Puglisi}
\affiliation{Institute for Complex Systems - CNR, P.le Aldo Moro 2, 00185, Rome, Italy}
\affiliation{Department of Physics, University of Rome Sapienza, P.le Aldo Moro 2, 00185, Rome, Italy}
\affiliation{INFN, University of Rome Tor Vergata, Via della Ricerca Scientiica 1, 00133, Rome, Italy}

\author{A. Sarracino}
\affiliation{Department of Engineering, University of Campania ``Luigi Vanvitelli'',  81031, Aversa (Caserta), Italy}
            
\begin{abstract}
  We investigate how the kinetic energy acquired by a dense granular
  system driven by an external vibration depends on the input energy.
  Our focus is on the dependence of the granular behavior on two main
  parameters: frequency and vibration amplitude. We find that there
  exists an optimal forcing frequency, at which the system reaches the
  maximal kinetic energy: if the input energy is increased beyond such
  a threshold, the system dissipates more and more energy and recovers
  a colder and more viscous state. Quite surprisingly, the
  nonmonotonic behavior is found for vibration amplitudes which are
  sufficiently small to keep the system always in contact with the
  driving oscillating plate. Studying dissipative properties of the
  system, we unveil a striking difference between this nonmonotonic
  behavior and a standard resonance mechanism. This feature is also
  observed at the microscopic scale of the single-grain dynamics and
  can be interpreted as an instance of negative specific heat. An
  analytically solvable model based on a generalized forced-damped
  oscillator well reproduces the observed phenomenology, illustrating
  the role of the competing effects of forcing and dissipation.
\end{abstract}

\maketitle

\section{Introduction}

The fascination of granular systems relies on
their rich phenomenology, which eludes standard
statistical physics~\cite{jeager}. Examples of surprising behaviors
were reported in experiments and numerical simulations:
Brazil-nut effect~\cite{Breu2003}, ratchet
effect~\cite{derMeer2004,gnoli2013}, spontaneous segregation of
mixtures~\cite{schroter2006}, nonequilibrium phase
transitions~\cite{garzo2011}, anomalous thermal
convection~\cite{gallas1992,pontuale2016}, Kovacs-like memory
effect~\cite{prados2014}, see also~\cite{aranson2006}. These features
are ascribed to the dissipative nature of these systems, which
generally cannot be treated via the introduction of effective
parameters~\cite{puglisi2017}. A fundamental open question is how the
dissipation mechanisms relevant for different behaviors are related to
the external energy injection.  These mechanisms involve several
scales, from particle-particle collisions to the interaction with
boundaries.

The relation between system kinetic energy $K$ and input energy $S$
involves the nonequilibrium response beyond the linear regime.  In
general, nonmonotonic behaviors have been observed in different
models, from driven tracers in crowding environments (negative
differential
mobility)~\cite{jack,sellitto,leitmann,baerts,benichou,sarracino}, to
systems showing negative specific heat due to long-range
interactions~\cite{lynden,barre,ramirez,staniscia} or to the presence
of baths at different temperatures~\cite{zia,bisquert}.  Similar
effects have been also observed in force-free cooling granular gases
of aggregating particles~\cite{brilliantov2}, where the granular
temperature can grow while the system energy decreases due
to dissipative collisions.

These behaviors affect fluidization properties of the medium, with
important consequences in industrial applications~\cite{coussot2005},
where usually energy is fed via mechanical vibrations with frequency
$f$ and amplitude $A$.  In some cases, a relation $K\sim S^\alpha$,
with $S\sim (Af)^2$, has been
derived~\cite{Luding1994,Kumaran1998,McNamara1998,Zivkovic2011}.
Experimental studies focused on the specific role of the forcing
mechanisms, such as vibration amplitude, frequency or
velocity~\cite{windows2015,windows2,lasta,gnoli2018,vidal2018,astrom2000,urb2004,DB05,capozza,johnsonjia2005,sellerio,griffa2013,wortel2014,giaccoprl2015,
  corwin2015,maksephyschemestry,dijk,gnoli2}.  In particular, an
optimal frequency for energy transfer is found when the system is in a
bouncing-bed state and $A$ and $f$ are varied keeping $S$
fixed~\cite{windows2015,windows2}. In this state the granular medium
detaches from the driving plate and a resonant behavior is achieved
via a synchronization between plate vibration and bed bouncing.

In this paper we consider vibration amplitudes small enough to keep
the granular system always in contact with the oscillating plate.  Our
main result is that, even in this regime, if the input energy fed into
the system is increased by increasing $f$ (keeping $A$ fixed), a
nonmonotonic behavior is observed in both the kinetic energy of a
driven vane immersed in the medium and in the kinetic energy of the
granular medium itself. We show that there is an optimal frequency
where the system reaches a maximum kinetic energy.  Differently from
previous results~\cite{windows2015,windows2}, our system is not in a
bouncing-bed or resonant state, and the granular internal energy is
nonmonotonic with $f$, even if the energy input always increases. Our
results can be interpreted as an example of negative specific heat,
shedding light on the findings reported in~\cite{brilliantov2}: Here,
we investigate the complementary process, where the granular
temperature decreases when the input energy increases.

The paper is structured as follows. In Sec.~\ref{sec:setup} we
describe the experimental setup and the numerical model.  In
Sec.~\ref{sec:optimal} we discuss the nonmonotonic behavior of the
system energy as a function of the input energy, which represents the
main result of the paper. Sec.~\ref{sec:velocity} is devoted to the
study of the vane angular velocity, which shows a similar nonmonotonic
behavior, related to the viscous properties of the granular medium.
In Sec.~\ref{sec:single} we present the analysis of the single-particle
dynamics, which provides further details on the physical mechanisms
underlying the observed phenomenology. The proposed theoretical model
of a generalized driven-damped oscillator is described in
Sec.~\ref{sec:model}. In Sec.~\ref{sec:comparison} we compare our
findings with previous results, pointing out the main
differences. Finally, in Sec.~\ref{sec:conclusion} we draw some
conclusions on the effects and mechanisms observed in our system.
Appendix~\ref{app} provides details on the numerical model.

\section{Experimental setup and numerical model}
\label{sec:setup}

\begin{figure}[!tb]
  \includegraphics[width=0.9\columnwidth]{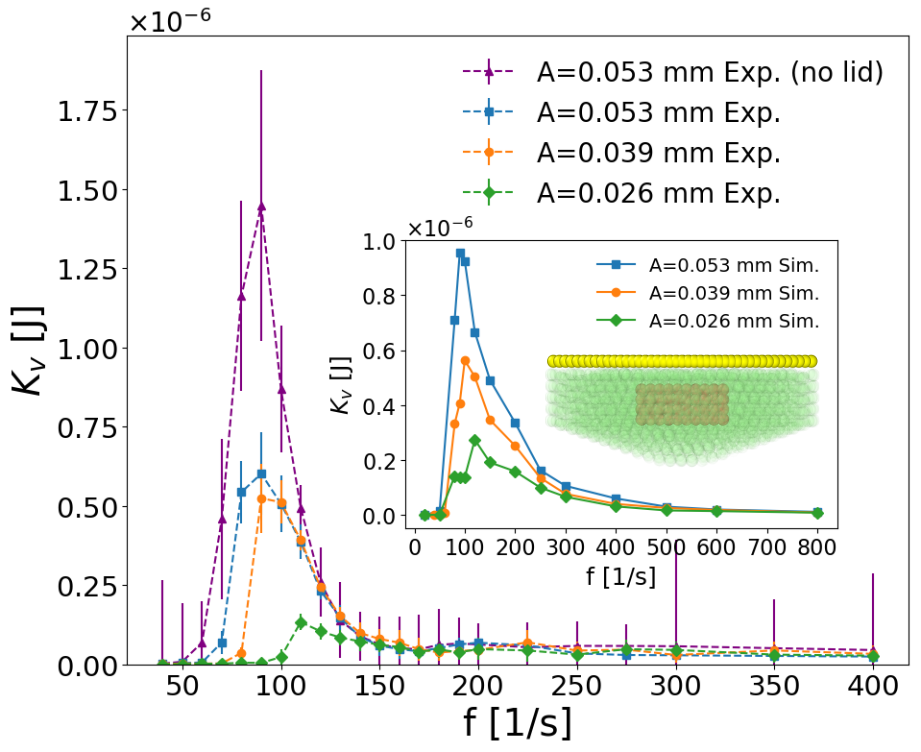}
\caption{Vane kinetic energy $K_v$ in experiments as a
    function of $f$, for different values of $A$. Error bars are $\pm 1$ standard deviation. Inset: $K_v$ measured in
    numerical simulations.  Sketch of the system setup: green spheres
  are granular particles, red spheres represent the vane, and yellow
  ones the lid. Numerical parameters are $k_n=12 \times 10^7$ Pa and
  $\gamma_n=2.9\times 10^7$ (ms)$^{-1}$.}
\label{fig00}
\end{figure}

Our experimental
setup is illustrated in Fig.~\ref{fig00} (see also
Ref.~\cite{gnoli2018}).  The granular medium is made of $N=2600$ steel
spheres (diameter $d=4$ mm, mass $m=0.27$ g), contained in a
conical-shaped floor cylinder (diameter 90 mm, minimum height 28.5 mm,
maximum height 47.5 mm), which enhances the energy transfer from
vertical to horizontal directions. A rigid lid (mass $M_{top}=218$ g)
covers the system, to confine particles and to
  allow for a spectral analysis of the system oscillations.  A vertical vibration is imposed to the coordinate of the container bottom $z_p(t)$
\begin{equation}
z_p(t)=A \sin(2\pi f t).
\label{zmax}
\end{equation}
A Plexiglas vane (height 15 mm, width 6 mm, length 35 mm) is suspended
in the system.  The vane can only rotate around the vertical axis,
subjected to a constant torque $\mathcal{T}=6 \times 10^{-3}$ Nm.

The system is simulated by LAMMPS package~\cite{LAMMPS}.
The interaction among grains is described via the nonlinear
Hertz-Mindlin model~\cite{brilliantov,silbert}, see Appendix~\ref{app}
for details. The relevant parameters are the stiffness of the
nonlinear contact $k_n$ and the viscous damping coefficient
$\gamma_n$, accounting for dissipative interactions.  The geometry of
the system and numerical parameters are chosen to reproduce our
experimental setup: The lid is made of 1773 granular particles
glued together and the rotating vane is made of $4\times 10$
particles glued together and overlapping by half a radius.

\section{Optimal forcing frequency and role of dissipation}
\label{sec:optimal}

In the main panel of Fig.~\ref{fig00} we report the vane kinetic
energy contribution due to fluctuations, $K_v=I [\langle
  \Omega^2\rangle-\langle \Omega\rangle^2]/2$~\cite{kv}, where $I=353$
g mm$^2$ is the momentum of inertia and $\Omega$ the angular velocity,
measured in experiments as a function of $f$ for different values of
$A$. We also show a case with no lid to demonstrate the robustness of
the observed behavior. In the inset we report results of numerical
simulations showing that the model well reproduces the nonmonotonic
behavior of the real system.
In the following, we will be mainly interested in the total kinetic
energy of the granular medium and in the dynamics of single grains,
and therefore we will remove the vane in some simulations of the numerical
model~\cite{footnote1}. Numerical simulations allow us to investigate
granular kinetic energy and single-grain motion, exploring a wider
range of vibration frequencies, $f\in(0,1000]$ Hz, with respect to the
  experiments.

As shown in Fig.~\ref{fig00}, the vane kinetic energy is a
nonmonotonic function of $f$ (at fixed $A$): $K_v$ grows abruptly from
zero to a finite value at a frequency threshold $f_1$, related to the
detachment condition~\cite{capozza}, $2\pi f_1=\sqrt{g/A}$.  Then,
after a maximum, $K_v$ starts to decrease, signaling that the granular
medium leaves the state of maximal fluidization, hindering the motion
of the vane by an increased effective viscosity.  One can define a
so-called friction recovery frequency, $f_2 > f_1$, at which the
system kinetic energy decays to about $1/2$ of its maximum value.
This behavior is related to dissipation mechanisms of the granular
medium that depend on the viscoelastic properties of the material.
\begin{figure}[tb!]
  \includegraphics[width=0.8\columnwidth]{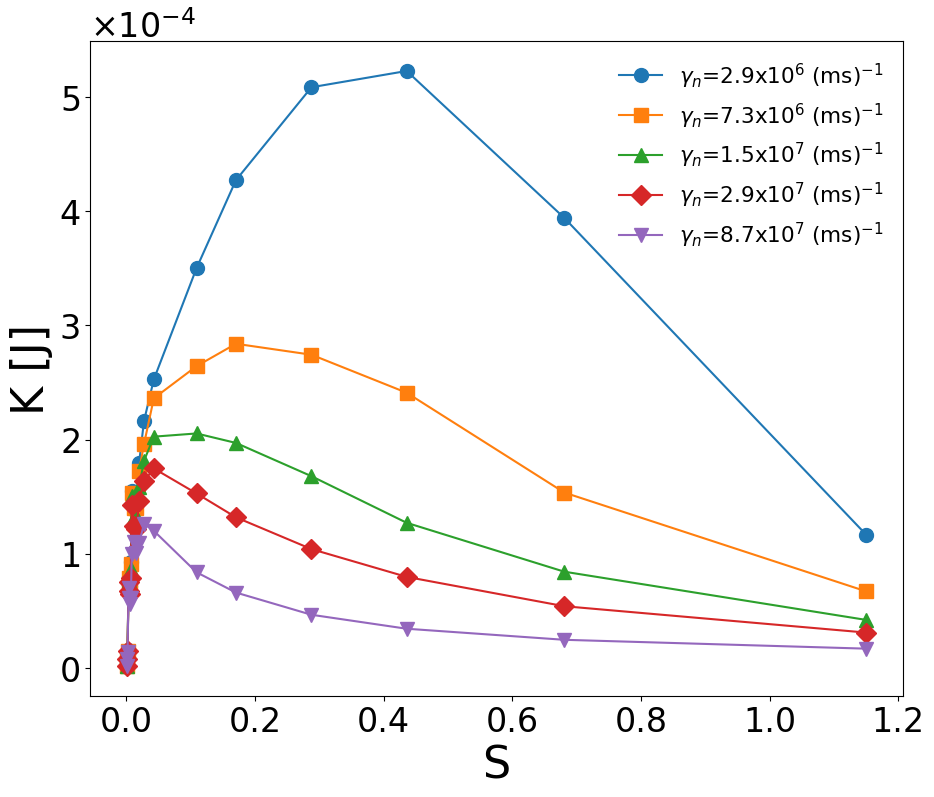}
\caption{Total kinetic energy $K$ of the granular medium versus $S$,
  varying $f$ at fixed $A$, for several values of $\gamma_n$, with
  $k_n=6.1\times 10^7$ Pa and $A=0.026$ mm.
  \label{fig:KeVariogN}}
\end{figure}

\begin{figure}[t!]
\includegraphics[width=0.35\textwidth]{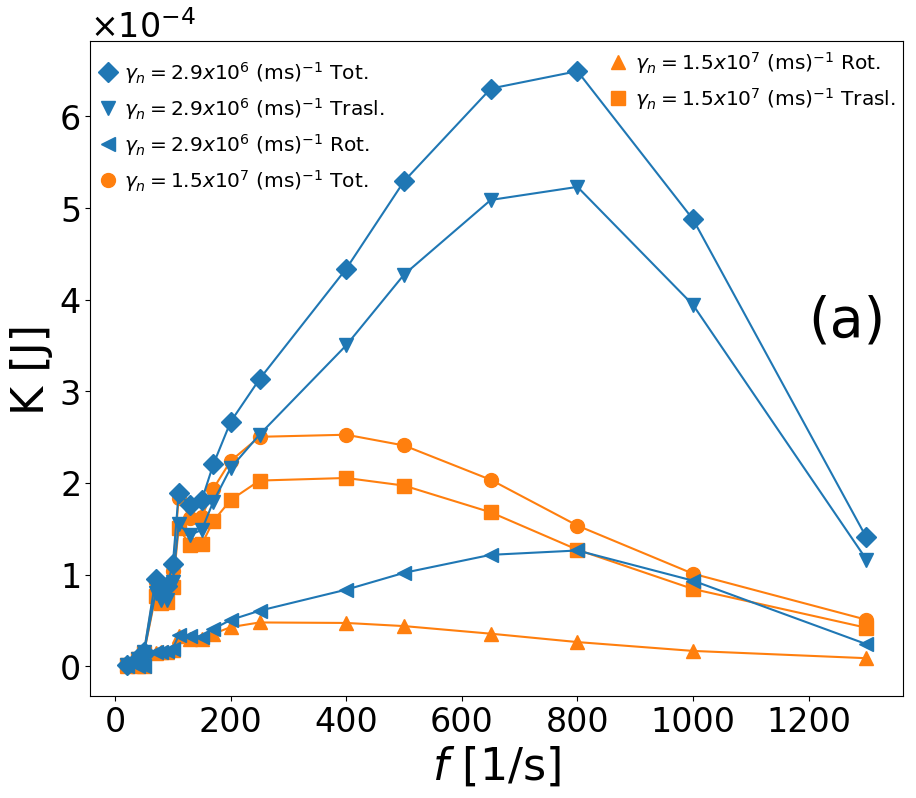}
\includegraphics[width=0.35\textwidth]{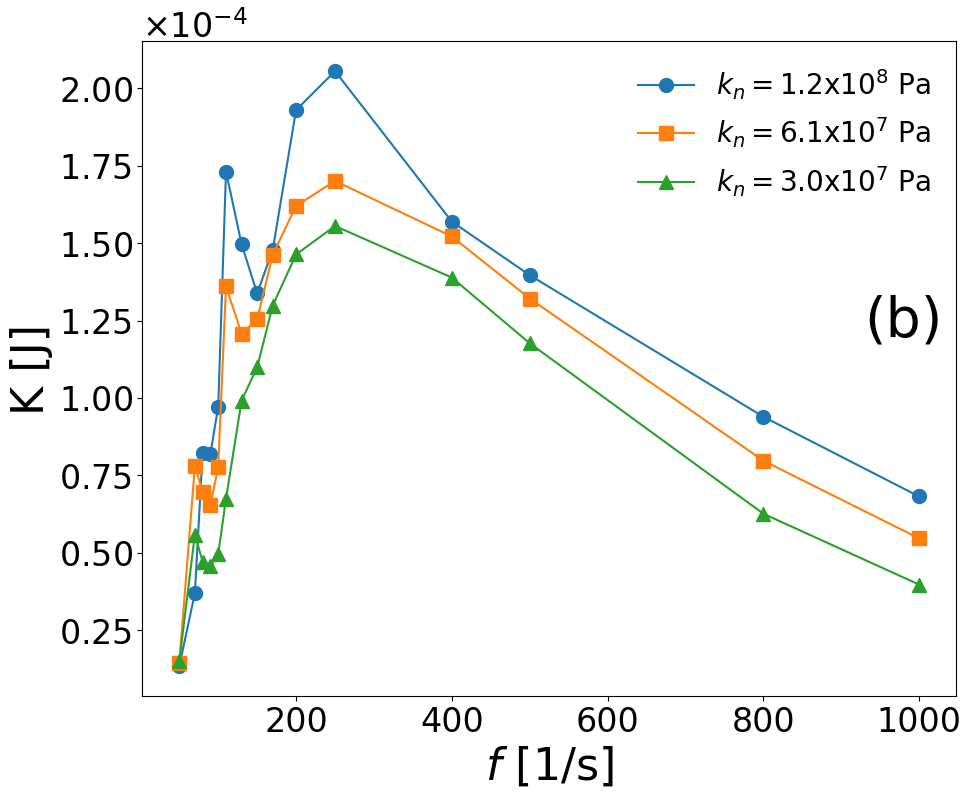}
\caption{Mean translational kinetic energy of the whole system $K$
  versus the driving frequency $f$ at fixed $A=0.026$ mm (simulations
  without the vane). (a) Comparison between the translational kinetic
  energy, the rotational one and the total (translational plus
  rotational). We show two curves reported in Fig. \ref{fig:KeVariogN}
  as a function of $f$, instead of $S$. The rotational degrees of
  freedom follow the same nonmonotonic behavior but with smaller
  absolute values, so that their contribution does not affect the
  qualitative behavior of the total kinetic energy.  (b) Study of the
  mean kinetic energy for different values of nonlinear stiffness:
  $k_n$ is varied at fixed $\gamma_n=2.9 \times 10^{7} $
  (ms)$^{-1}$. The three curves have the same shape (the peak position
  don't change) but are vertically ordered by $k_n$. This implies a
  larger kinetic energy and therefore less dissipation in the system
  with higher stiffness.\label{fig:Ke}}
\end{figure}

To better characterize the intrinsic behavior of the granular medium
and its response to the external forcing, we focus on the total
translational kinetic energy (the rotational kinetic energy being
negligible, see Fig. \ref{fig:Ke}a) $K=\sum_{i=1}^Nm\langle
\boldsymbol{v}_i^2\rangle/2$, where $ \boldsymbol{v}_i$ is the grain
velocity, in the absence of the suspended vane. Therefore, we
introduce the adimensional quantity describing the energy input,
$S=(2\pi f)^2 A^2/(gd)$.

In Fig.~\ref{fig:KeVariogN} we report $K(S)$, varying $f$ at fixed
$A$, for several values of $\gamma_n$. The nonmonotonic behavior shows
that the granular system reaches the maximal kinetic temperature at an
optimal value of the input energy. When the input energy exceeds a
certain threshold, related to $f_2$, the system cools down because the
dissipation effects increase.  A key result is represented by the
maximum position of the kinetic energy for different values of viscous
damping coefficient $\gamma_n$. We find that the peak shifts to the
left when the dissipation in the system is increased. It is
interesting to note that this shift doesn't occur if we vary $k_n$
signaling that the friction-recovery mechanism is not governed by the
elastic contribution of the interaction (see Fig. 3(b)). Therefore the
system can transfer more energy to the grain motion when it is
vibrated at an optimal frequency. If the frequency increases, the
overall external energy injected is larger but the dissipation
mechanisms become dominant and the granular kinetic temperature
decreases.  This behavior can be interpreted as an instance of
negative specific heat~\cite{zia,brilliantov2}, occurring due to the
subtle interplay between forcing and dissipation.

\begin{figure}[tb!]
\includegraphics[width=0.35\textwidth]{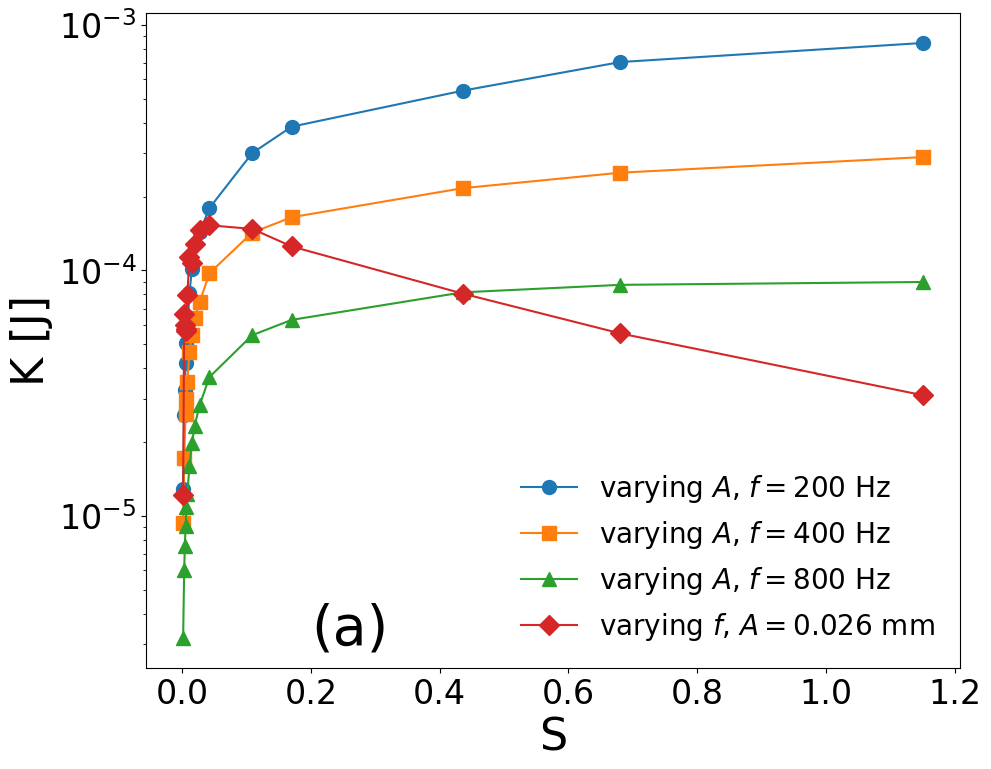}
\includegraphics[width=0.35\textwidth]{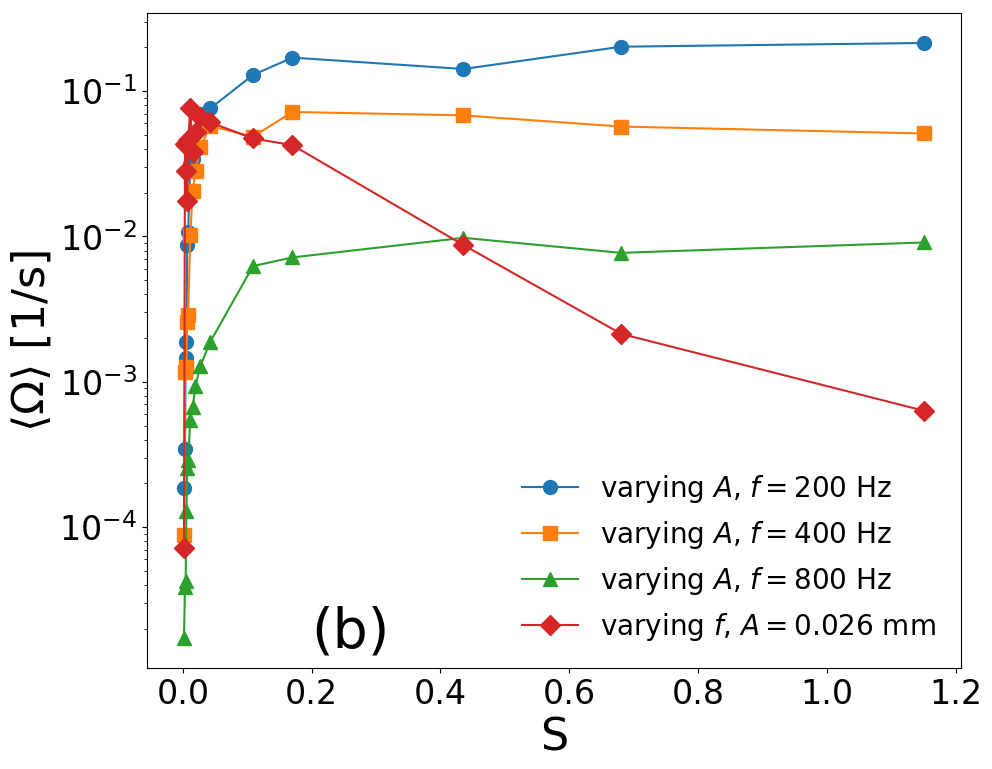}

\caption{Total kinetic energy $K$ of the grains (panel (a)) and mean angular velocity $\langle \Omega\rangle$ of the blade  (panel (b)) versus $S$, varying $A$ for three fixed
  frequencies (200 Hz, 400 Hz and 800 Hz), and varying $f$ for one
  fixed amplitude ($A$=0.026 mm).  Parameters are $k_n=6.1 \times
  10^7$ Pa and $\gamma_n=2.9 \times 10^{7}$ (ms)$^{-1}$.\label{fig:VarioAmpKe}}
\end{figure}

%Finally, we investigate the role played by changing the shaking amplitude at fixedfrequency. 
Conversely, if we increase $S$ by increasing $A$ at fixed $f$, we find
a monotonic behavior. Namely, dissipative mechanisms are not activated
and the kinetic energy keeps growing with the input energy as shown in
Fig.~\ref{fig:VarioAmpKe}. These features are a consequence of the
permanent contact with the driving plate, which makes dissipation
dominating at high frequencies.  This is a novel phenomenon different
from the resonant behavior in the bouncing-bed state clearly described
in Ref.~\cite{windows2015,windows2}, where the optimal frequency is
an increasing function of dissipation. Striking differences are also
provided by the spectral analysis of the top plate oscillations which
indicates that, in our system, the energy transfer is not maximized in
the most coherent states (see Sec.~\ref{spectra}).
 
\section{Vane average velocity}
\label{sec:velocity}

Another crucial quantity in our experimental/numerical setup is the
average angular velocity $\langle \Omega\rangle$ of the vane in
the steady state. While the variance is related to the kinetic energy
of the system, the mean value of $\Omega$ is proportional to the
inverse of the viscosity acting on the tracer during its motion.  In
the following we present the numerical study of $\langle \Omega
\rangle$ and $K$ with particular attention on the role played by the
parameters $k_n$ and $\gamma_n$ of the HM model in the
characterization of its behavior. For an experimental study of this
observable in the same setup see \cite{gnoli2018}.

\begin{figure}[t!]
\includegraphics[width=0.3\textwidth]{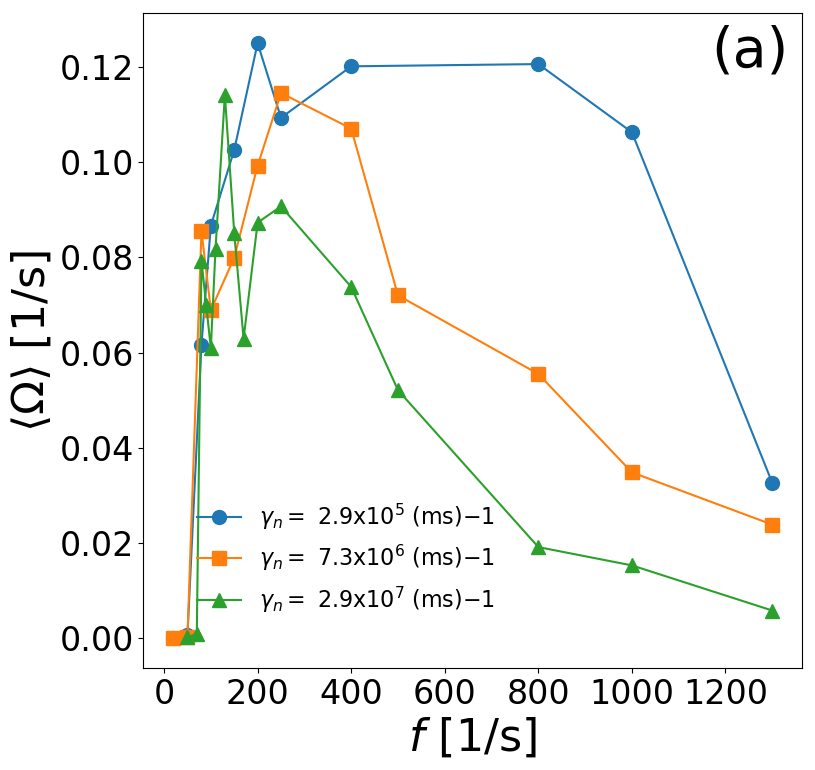}
\includegraphics[width=0.3\textwidth]{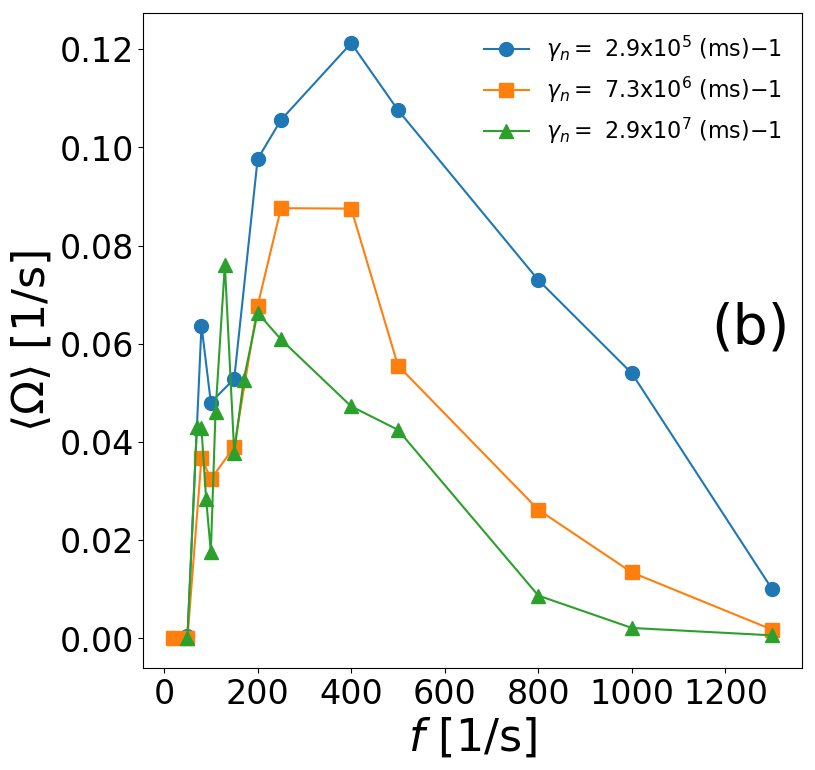}
\includegraphics[width=0.3\textwidth]{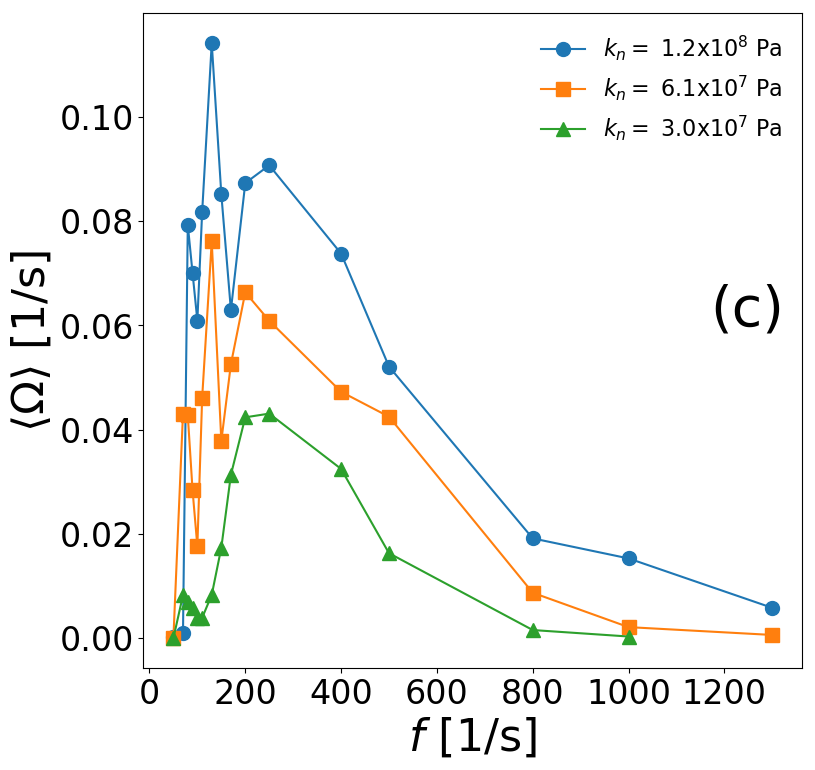}
\caption{Mean angular velocity of the blade $\langle \Omega\rangle$ versus the
  driving frequency $f$. (a) $\gamma_n$ is varied with fixed $k_n=1.5
  \times 10^{8}$ Pa. (b) $\gamma_n$ with fixed $k_n=6.1 \times 10^{7}$
  Pa. (c) $k_n$ is varied with fixed $\gamma_n=2.9 \times
  10^{7}$ (ms)$^{-1}$. \label{fig:KnFix}}
\end{figure}

In Fig. \ref{fig:VarioAmpKe}b  we show $\langle \Omega \rangle$  as a function of
$S$ by increasing $f$ at fixed $A$ and vice-versa.  We find a behavior very similar to the one of $K$ reported in the previous section, the only qualitative difference is that $\langle\Omega\rangle$
versus $A$ reaches a constant value, while $K$ versus $A$ keeps growing also at larger amplitudes (compare with panel (a) of the same figure).
To better understand the role of the interaction parameters we concentrate on $\langle\Omega\rangle (f)$ at constant amplitude.
In Figs. \ref{fig:KnFix}(a) and (b) we vary $\gamma_n$ at fixed $k_n$, while
in Fig. \ref{fig:KnFix}(c) we vary $k_n$ at fixed $\gamma_n$.  We see
that the parameter $k_n$ changes the typical shape of the curves near
$f_1$: we have a sudden peak when $k_n$ is high and a smoother
behavior when it is small. In Fig. \ref{fig:KnFix}(b) we observe that at
lower $k_n$, reducing $\gamma_n$ clearly slows down the friction
recovery. On the other hand, at higher $k_n$ (Fig. \ref{fig:KnFix}(a))
we see the same effect but without a significant change in the height
of the peak. This can be interpreted intuitively observing that a
higher stiffness reduces the effect of the viscous damping during the
collision. In Fig. \ref{fig:KnFix}(c) we see that also an increase of
$k_n$ implies a growth of the recovery frequency, but this effect is
less pronounced with respect to the one obtained decreasing
$\gamma_n$.
We can conclude that $f_2$
is mostly ruled by the dissipation through $\gamma_n$, in agreement
with previous experimental results~\cite{gnoli2018}. 

To sum up, a comparison between Figs. \ref{fig:KnFix}(a), \ref{fig:KnFix}(b)  and \ref{fig:Ke}(a)
shows that the behavior of both $\langle \Omega\rangle$ (a rheological
response) and $K$ (a thermodynamic observable) as a function of $f$ is
similarly affected by the change of $\gamma_n$. On the contrary,
from Figs. \ref{fig:KnFix}(c) and \ref{fig:Ke}(b), it is clear that
changing $k_n$ affects $\langle \Omega\rangle$ more than $K$.

%\subsection{Varying the amplitude at constant frequency}

\section{Analysis of the single-particle dynamics}
\label{sec:single}

The macroscopic features above described can be related to microscopic
properties by investigating the single-grain dynamics.  Typically, in
dense systems, the diffusive motion of a single particle is hindered
by the presence of many surrounding particles, inducing a cage
effect~\cite{scalliet,plati2019,giacco2017}. Focusing on the Mean
Squared Displacement (MSD) of a particle, one observes that, after a
ballistic motion at short times, a plateau develops, signaling that
the particle is trapped inside a cage. Then, at longer times, the
particle manages to explore a larger region of the system and its MSD
grows linearly in time.  The relevant quantities are the cage size
$S_c$ and the trapping time inside a cage $\tau_c$, both estimated
from the first point of the MSD after the ballistic regime.
\begin{figure}[tb!]
\includegraphics[width=0.8\columnwidth]{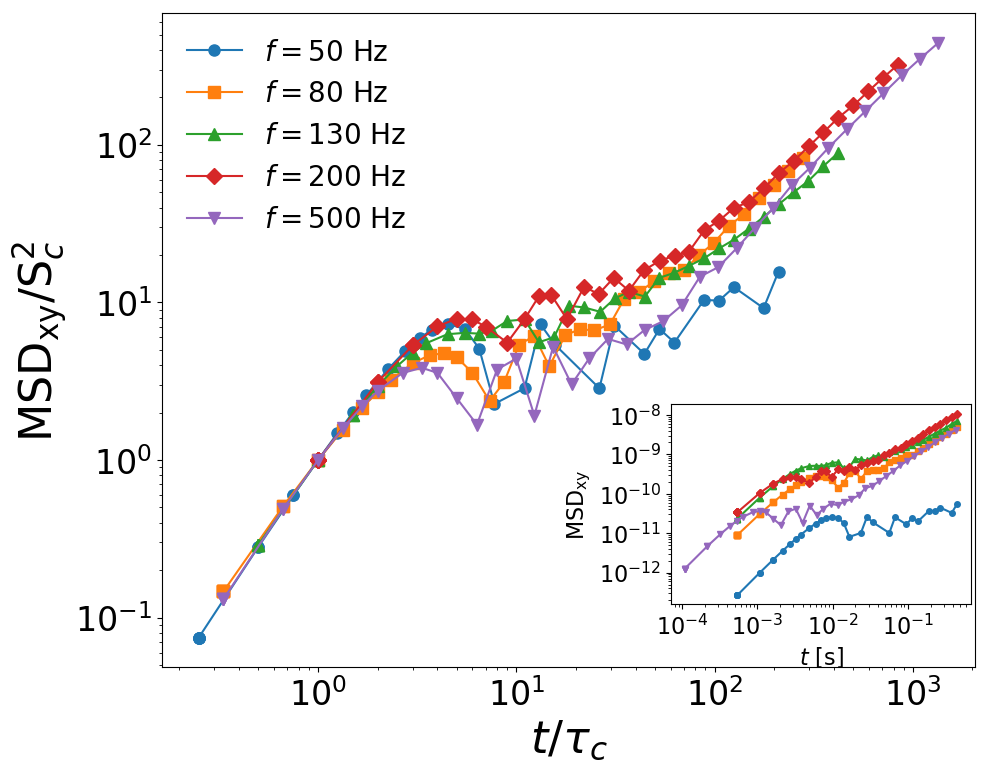}
\caption{Rescaled single-particle MSD in the $XY$ plane versus
  rescaled time, for different values of the driving frequency. Inset:
  single-particle MSD in the $XY$ plane versus time. Simulation
  parameters are $A=0.026$ mm, $k_n=6.1 \times 10^7$ Pa and
  $\gamma_n=2.9 \times 10^{7}$ (ms)$^{-1}$.
  \label{fig:MsdFreq}}
\end{figure}
\begin{figure}[tb!]
  \includegraphics[width=0.8\columnwidth]{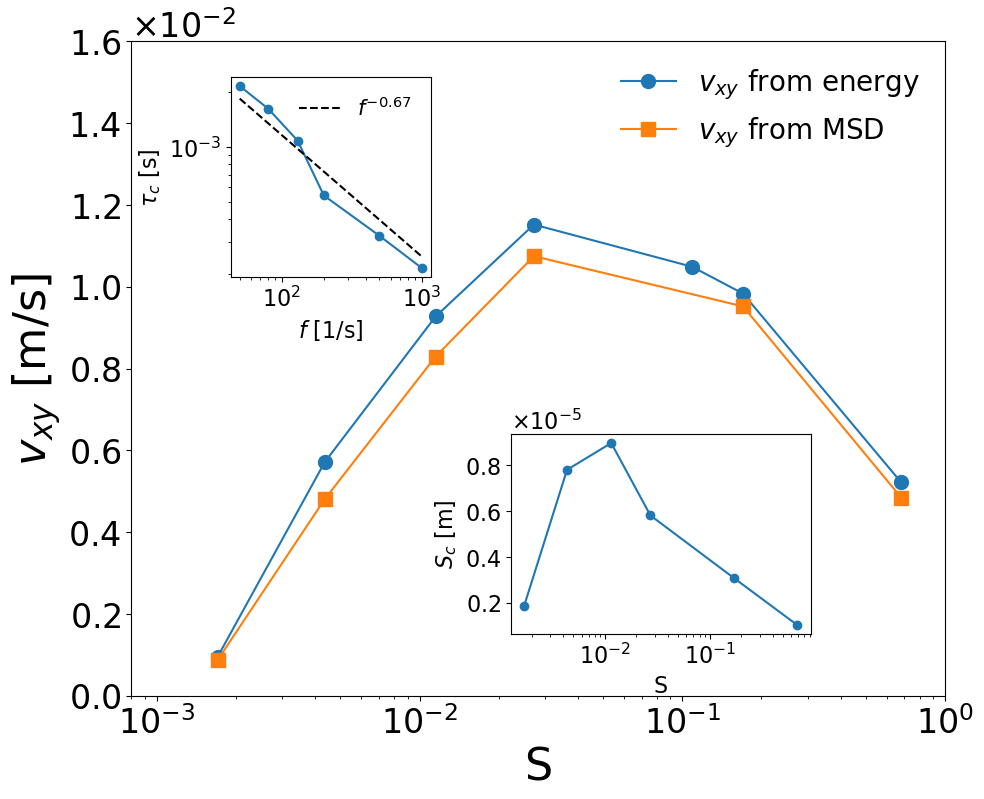}
\caption{Comparison between the average particle speed in the $XY$
  plane obtained from the mean kinetic energy of the whole system and from the ratio between
  the cage size $S_c$ and the trapping time $\tau_c$.  Bottom inset:
  $S_c$ as a function of input energy $S$, varying $f$ at fixed
  $A$. Top Inset: $\tau_c$ as a function of $f$. Same parameters of
  Fig.~\ref{fig:MsdFreq}.}\label{fig:sizecage}
\end{figure}
This picture is fully supported by Fig.~\ref{fig:MsdFreq}, where we
show the MSD (averaged over about 20 particles randomly chosen in
the system) in the horizontal $XY$ plane, the plane relevant
for the vane's motion.  Rescaling time by the trapping time $\tau_c$
and the MSD by the squared cage size, we find that curves collapse.

\begin{figure}[t!]
\includegraphics[width=0.35\textwidth]{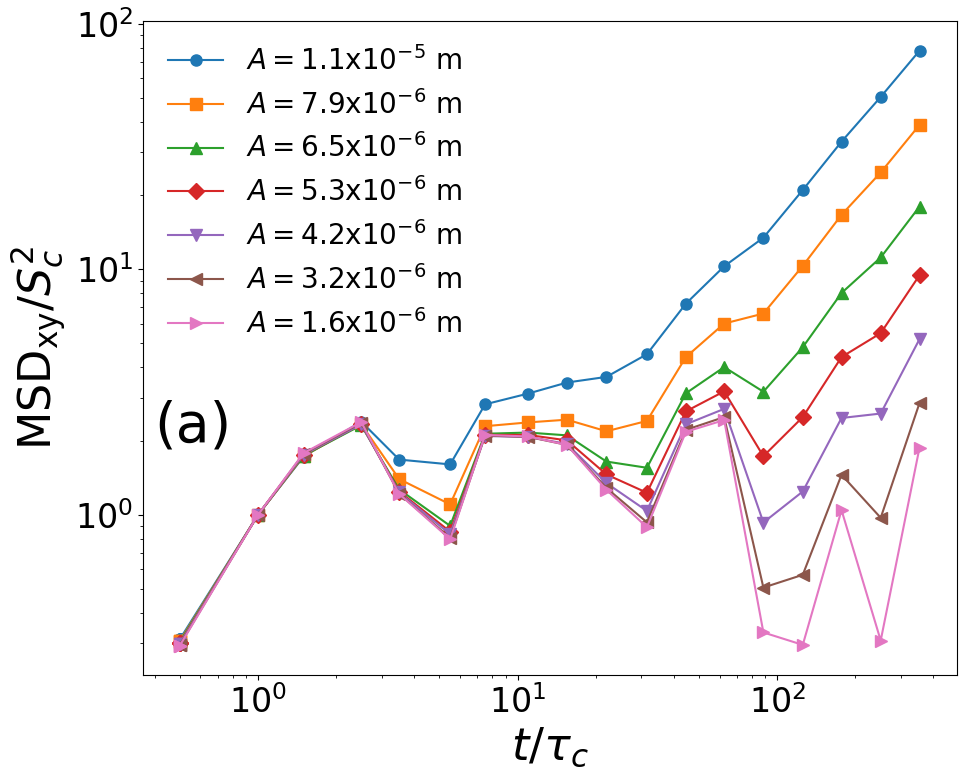}
\includegraphics[width=0.35\textwidth]{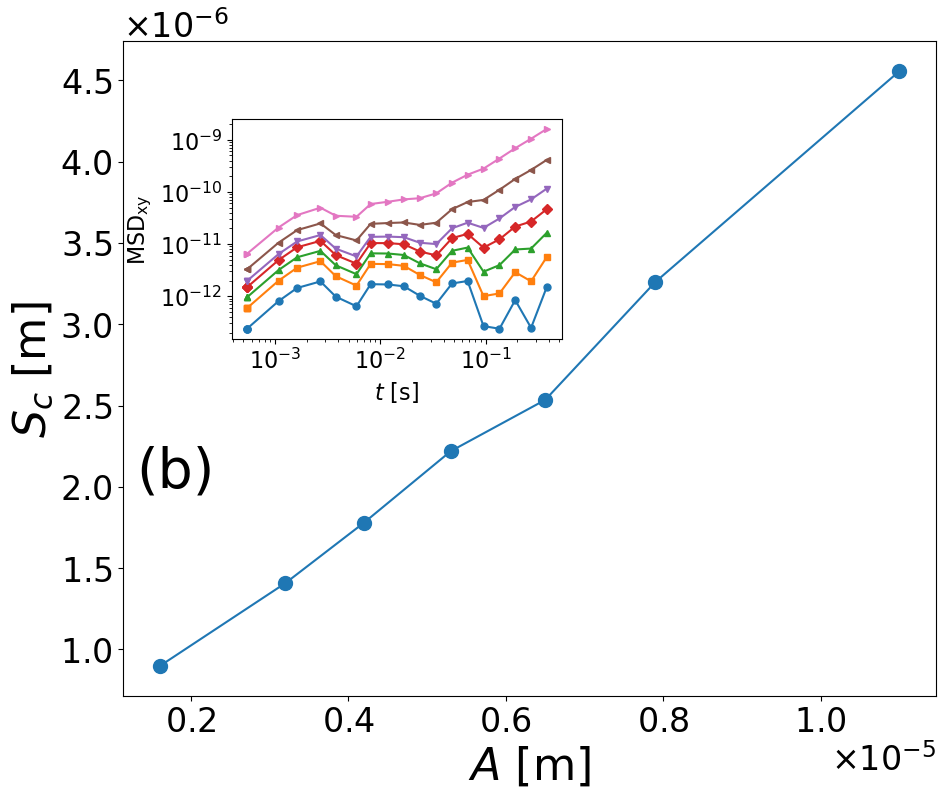}
\caption{A: Single-particle MSD in the $XY$ plane rescaled by
  $S_c^2$ as a function of the
  rescaled time $t/\tau_c$. Different curves correspond to different driving amplitudes
  while the frequency is fixed at $f=200$ Hz. B: Size of the cage
  $S_c$ versus the driving amplitude $A$. The inset shows the MSD not
  rescaled. Simulations with $k_n=6.1 \times 10^{7}$ Pa and
  $\gamma_n=2.9 \times 10^{7} $(ms)$^{-1}$. \label{fig:MsdAmp}}
\end{figure}

The nonmonotonic behaviors described above for $K_v$ and $K$, can be
rationalized by the study of $S_c$ and $\tau_c$, as reported in the
insets of Fig.~\ref{fig:sizecage}.  We find that $\tau_c$ is a
decreasing function of the frequency, $\tau_c \sim f^{-0.67}$, whereas
$S_c$ is a nonmonotonic function of $S$, varying $f$ at fixed $A$,
with a maximum at $S\simeq 10^{-2}$.  At low frequencies the cage size
tends to be very small but the particles need a long time to be
trapped. Conversely, at high frequencies cages are still small but the
particles are trapped in a short time. Moreover at low frequencies,
since the system is weakly perturbed by the input energy, a particle
is able to explore larger regions for increasing $S$. However, for
larger $S$ (increased through $f$), collisions become more frequent
and the explored cage region decreases accordingly.  If we estimate
the average particle speed $v_{\textrm{XY}}$ on the $XY$ plane as
$v_{\textrm{XY}}\sim S_c/\tau_c$, we find a nonmonotonic behavior, in
agreement with the behavior of $K$ (Fig.~\ref{fig:sizecage}). This
analysis shows that the single-particle dynamics reflects the same
phenomenology occurring at the macroscopic scale.

Moreover, the nonmonotonic behavior of $S_c$ could be related to a
change in the effective number of degrees of freedom in the system as
the input energy is varied, in agreement with the explanation of
negative specific heat suggested in Ref.~\cite{brilliantov2} for
cooling granular gases of aggregating particles. In that case,
although the total energy of the system decreases due to dissipative
collisions, the total number of clusters can decrease faster due to
agglomeration, producing an increase of energy per particle.

We show in Fig. \ref{fig:MsdAmp} the rescaled MSD on the xy-plane
averaged over 20 particles and the relative cage size $S_c$ for
different values of $A$. Here we see that the collapse is very good up
to time $t/\tau_c\sim 1$ (so the trapping time $\tau_c$ does not vary
with $A$) while $S_c$ (and consequently the speed $v_{xy}=S_c/\tau_c$)
grows monotonically with the driving amplitude. These last results
show that the same phenomenology as a function of $A$ at fixed
frequency is present both at the macroscopic scale
($\langle\Omega\rangle$ and $K$) and at the single-particle one
($v_{xy}$).

\section{The generalized driven-damped oscillator: a single-particle perspective}
\label{sec:model}

Here we present a generalized model of a driven-damped oscillator that
reproduces qualitatively the phenomenology of $K$ studied in the
simulations. In particular, our model shows: i) a nonmonotonic
behavior of the energy as a function of the driving frequency $f$ with a
maximum in $f^*$; ii) an increasing behavior of the energy as function
of the driving amplitude $A$; iii) the shift to the left of the
frequency $f^*$ when the dissipation in the system is increased.

The model is obtained starting from an equation of motion for a
generic particle in the system and then assuming that the energy of
the whole system follows the same behavior of that of the single
particle. This assumption is consistent with the previous analysis
that shows a common phenomenology of single-particle quantities and
macroscopic ones. In addition, the same assumption is done in PEPT
experiments where the time-averaged behavior of a single tracked
particle is considered representative of that of the global system
\cite{windows2015}.

\begin{figure}[t!]
\centering
\includegraphics[width=0.35\textwidth]{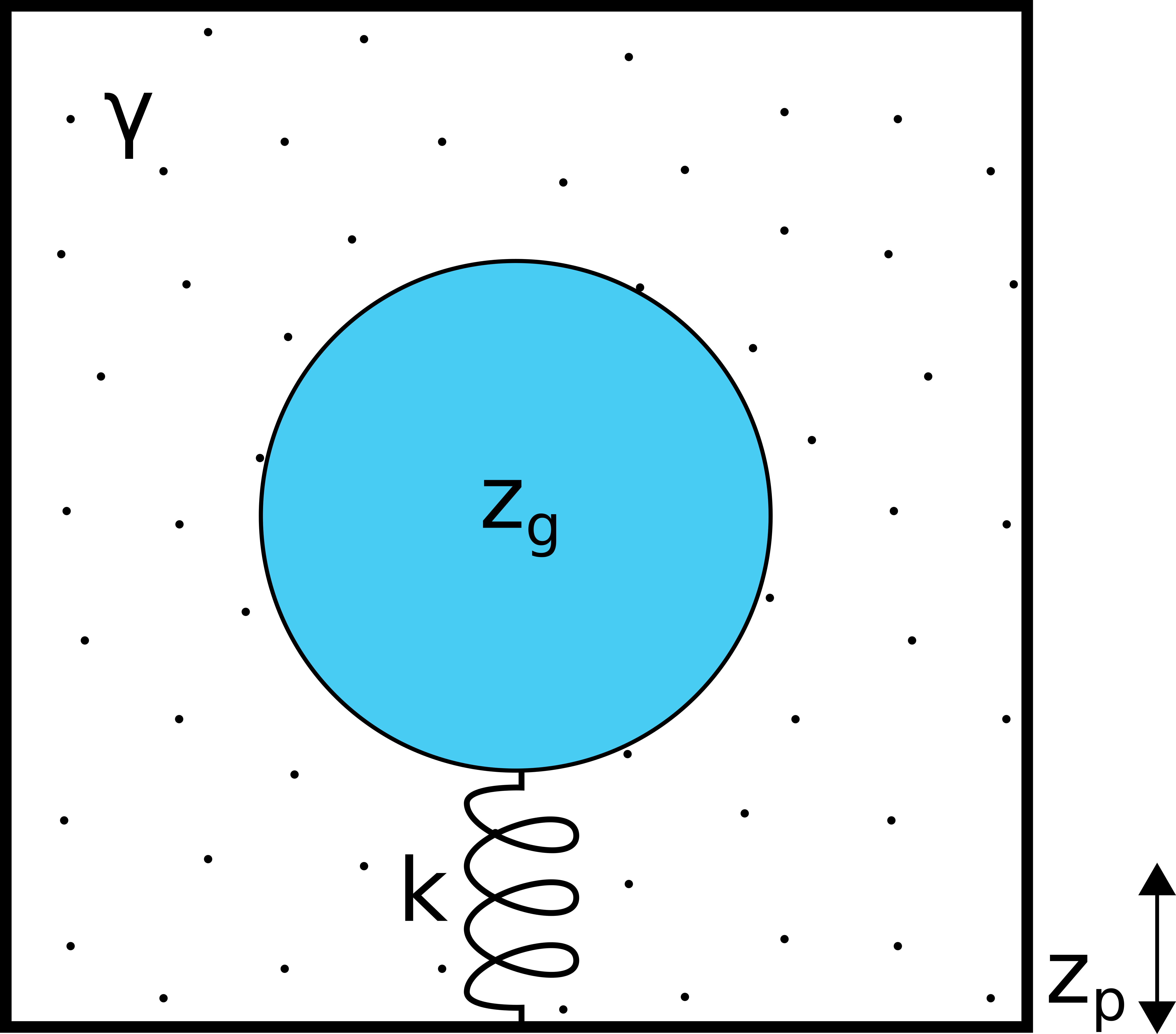}
\caption{Coarse-grained description of the single-particle dynamics. The springs $k$ and the viscosity $\gamma$ represent the cage that mediates the vibration of the shaker.} \label{fig:Sketch}
\end{figure}

\subsection{Newton equation for a caged single particle}

The granular system is very dense in all regimes of shaking, so the
short-time dynamics of the grains is expected to take place inside
their cages formed by the surrounding particles. This short-time
dynamics presents multiple relevant time-scales such as the one
represented by the integration time $dt$ ($\tau_1\sim 10^{-5}$ s),
that is a fraction of the duration of a single collision, and the one
fixed by the inverse of the driving frequency $\tau_d=1/f\sim
10^{-3}-10^{-1}$ s. A third one is represented by the mean time
between two collisions, reported in Fig. \ref{fig:sizecage} (but also
observed for the $XYZ$ diffusion), which spans values in the range
$\tau_c\sim 10^{-4}-10^{-3}$ s. These time-scales are thus ordered as
follows: $\tau_1 < \tau_c < \tau_d$ for all the cases presented in our
study.

In order to estimate the mean kinetic energy over many collisions, we
are interested in a time coarse-graining on a scale larger than
$\tau_c$. At the same time, we want to write a differential equation
for the coordinates of a particle in which the shaker dynamics appears
as an external driving. In view of these arguments, we concentrate on
time scales larger than $\tau_c$ but smaller than $\tau_d$. To make
our model as simple as possible we only consider the motion in the $z$
direction. Therefore we consider a single particle of mass $m$
confined in a one dimensional cage that consists in one spring of
stiffness $k$ with a resting length $l_0$ connected with the bottom of
a vibrating box. The latter really represents the
experimental/numerical container that oscillates following
$z_p(t)=A\cos(2\pi f t)$. The fact that the cage is actually made of
fast vibrating particles also induces an effective viscosity with
coefficient $\gamma$.

Our simplified coarse-grained description is sketched in
Fig.~\ref{fig:Sketch}, where we refer to the coordinate of the grain as
$z_g$. We remind here that the coefficients of the HM model used in
the simulation ($k_n$ and $\gamma_n$) depend on the material
properties and act on the fast time scale $\tau_1$. The way in which
they are connected with the effective viscosity $\gamma$ and stiffness
$k$ is not trivial.  A reasonable value for $k$ can be estimated
considering that the modeled spring is actually made by a column of
grains. Every grain with radius $r$ and Young modulus $Y$ can be
thought of as a microscopic vertical spring with an elastic constant
$\tilde{k}$ given by $\tilde{k}=\pi Yr/2$ (for simplicity we are
considering the grains as cylinders instead of spheres). Now the
effective stiffness of the column results from the parallel of a mean
number $\bar{n}$ of such microscopic springs:
$k=\tilde{k}/\bar{n}$. For the parameters of our simulations and
fixing $\bar{n}=4.5$ we find that $k\sim \mathcal{O}(10^{5})$ N/m.
Regarding the effective
viscosity,  we will do an ansatz  in the following. Looking at Fig. \ref{fig:Sketch}, it is straightforward to
write for $z_g$:
\begin{equation}
 \ddot{z_g}(t)+4\pi f_s\dot{z_g}(t)+(2\pi f_k)^2\xi(t)=0, \label{EqOAsimp}
\end{equation}          
where $\xi(t)=z_g-z_p-l_0$, $2\pi f_s=\gamma/(2m)$ and $2\pi f_k = \sqrt{k/m} \sim \mathcal{O}(10^{4})$ s$^{-1}$.

Now we come to the crucial hypothesis of our approach, i.e. the ansatz
on $f_s$. Some previous studies \cite{gnoli2018,windows2015} suggest
that the dissipation of energy due to interparticle collisions
increases significantly when the driving frequency $f$ grows. This is
also visible in our data from Fig. \ref{fig:sizecage}, where we see
that $\tau_c$ becomes smaller for increasing $f$. Indeed, a reduction
of the time between collisions means a growth of the number of
dissipative events in the system (i.e. the collisions themselves). The
simplest way to insert this dependence of the internal dissipation on
the external driving is to take $f_s$ as an increasing function of
$f$: $f_s=a f^{\alpha}$ with $a,\alpha>0$. Bringing the variable $z_p(t)$
contained in $\xi(t)$ to the right-hand side of Eq. (\ref{EqOAsimp}) and
adding gravity, we obtain the following equation:
\begin{eqnarray}
&&  \ddot{z_g}(t)+4\pi a f^{\alpha}\dot{z_g}(t)+(2\pi f_k)^2 z_g(t) \nonumber \\
  &=&(2\pi f_k)^2A\cos(2\pi f t) + (2\pi f_k)^2 l_0 -g. \label{EqOAfull}
\end{eqnarray} 
This is the equation for a driven-damped harmonic oscillator with
characteristic frequency $f_k$, viscous constant $4\pi a f^{\alpha} $
and external driving $(2\pi f_k)^2A\cos(2\pi f t)$ that oscillates
around the equilibrium position $z_g^{eq}=l_0-g/(2\pi f_k)^2$.  The
stationary solution of Eq. (\ref{EqOAfull}) is:
\begin{eqnarray}
  z_g(t)&=&\frac{f_k^2A}{\sqrt{(f_k^2- f^2)^2+4a^2f^{2(\alpha+1)}}}\cos(2\pi f t-\phi)+z_g^{eq}, \nonumber \\
  \phi&=&\arctan\left( \frac{2af^{\alpha+1}}{f_k^2- f^2} \right).
\end{eqnarray}  
Deriving $z_g(t)$, taking the square and then integrating over a period $1/f$ we can find the mean quadratic velocity of the particle:
\begin{equation}
\langle\dot{z_g}^2\rangle=\frac{\frac{1}{2}A^{2} f_k^4 (2\pi f)^{2}}{  4a^2 f^{2(\alpha+1)}   +  \left(f_k^2- f^{2}\right)^{2}}. \label{SoluzSimp}
\end{equation} 

\begin{figure}
\includegraphics[width=0.4\textwidth,clip=true]{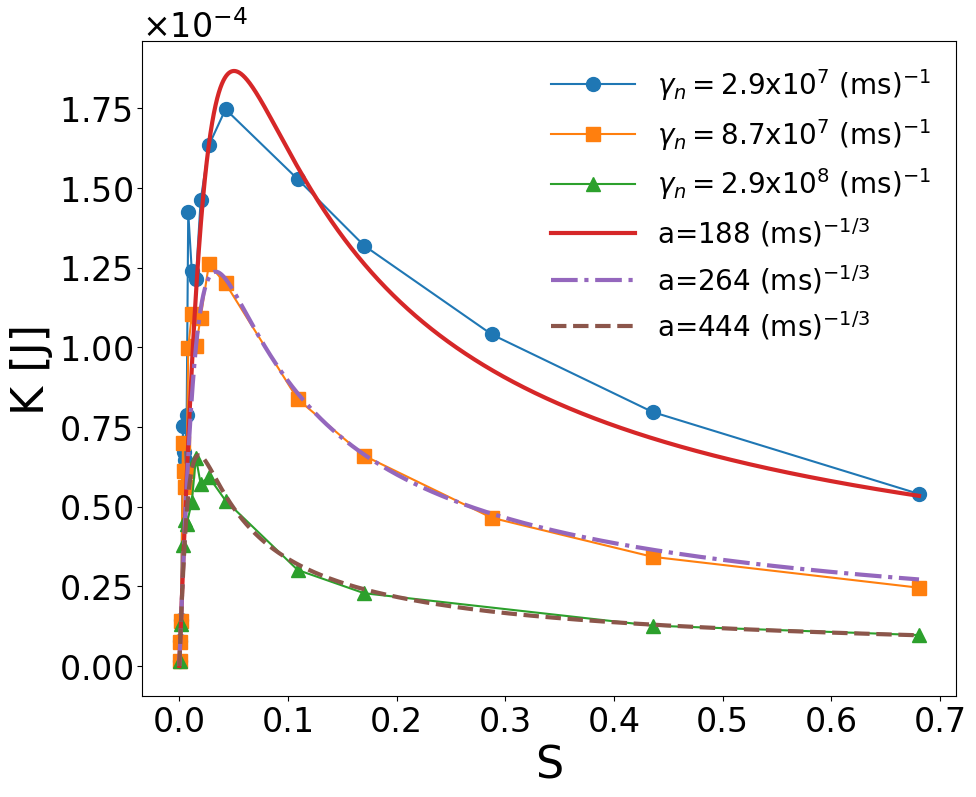}
\caption{Mean kinetic energy from Eq.~(\ref{SoluzSimp}) for different
  values of $a$, with $2\pi f_k=11905$ s$^{-1}$, $A=0.026$ mm and
  $m_{\textrm{eff}}=912.6$ g, compared with $K$ measured in numerical
  simulations  with $A=0.026$ mm and $k_n=6.1\times10^7$ Pa.
  \label{fig:modVSdat}}
\end{figure}

As showed in Fig.~\ref{fig:modVSdat} for the specific value
$\alpha=2/3\sim 0.67$ taken from the behavior of $\tau_c$ (but it holds in general for $\alpha>0$),
$\langle\dot{z_g}^2\rangle$ has a nonmonotonic behavior for $f<f_k$
and its maximum value shifts to the left as $a$ is increased.  In
Fig.~\ref{fig:modVSdat} we show the behavior of the mean total kinetic
energy $K=m_{\textrm{eff}}\langle\dot{z_g}^2\rangle/2$ as a function
of $S$, for several values of $a$.  From a fitting procedure, we found
$2\pi f_k=11905$ $s^{-1}$, which is of the order $\sqrt{k/m}$ if the
effective stiffness $k$ is estimated by considering series-parallel
microscopic elastic constants relative to the grain material, as above
illustrated.  The prefactor $m_{\textrm{eff}}$ represents an effective
mass and we find an optimal agreement with data for
$m_{\textrm{eff}}=912.6$ g, which is of the order of the total mass of
the system. Therefore, the analytical model predicts the general phenomenology of
the 3D granular system, with a good quantitative agreement for high
values of $\gamma_n$.

\subsection{Beyond a simple resonance}

Eq.~(\ref{SoluzSimp}) has a very simple form and its nonmonotonic
behavior can be explained in terms of competitive mechanisms. From
Eq. (\ref{SoluzSimp}) we see that the mean quadratic velocity of the
particle is proportional to the input energy of the shaker divided by
an adimensional factor: $\langle\dot{z_g}^2\rangle=V(f)/U(f,\alpha)$
where $V(f)=\frac{1}{2}A^{2}(2\pi f)^{2}$ is proportional to the
strength of the shaker $S=(2\pi f)^2A^2/(gd)$ while
$U(f,\alpha)=u_1+u_2$. Here $u_1(f,\alpha)=4a^2 f_k^{-4}
f^{2(\alpha+1)}$ and $u_2(f)=(1-f^2/f_k^2)^2$ can be considered the
two competitive terms if $f<f_k$. In this regime, $u_2$ is a
decreasing function of $f$ and enhances energy transfer, while $u_1$
(that contains the dissipation) increases with $f$ and therefore has
an opposite effect. In order to better understand the underlying mechanisms,
we remind that for an ordinary driven-damped oscillator (namely with a
viscous coefficient that does not depend on $f$) the mean quadratic
velocity is the same as Eq. (\ref{SoluzSimp}), with the only difference
that the dissipative term is proportional to the square of the
rescaled driving frequency: $u_1=u_1(f,0)$. From this point of view,
$1/u_2$ can be interpreted as an energy gain factor: It grows before the
resonant frequency ($f=f_k$) and then decreases. In the case of the
usual damped oscillator, therefore, the nonmonotonic behavior is
entirely explained by the nonmonotonic behavior of $u_2$ alone, and
in fact dissipation does not change the peak position but only
smoothes it. On the contrary, when $\alpha>0$ the nonmonotonic
behavior has a different origin, coming from the competition between
dissipation ($u_1$) and gain ($u_2$).  This can be rigorously
checked, deriving Eq. (\ref{SoluzSimp}) to find the condition for the
maximum:
\begin{equation}
f_k^4-f^4-4\alpha a^2 f^{2(\alpha+1)}=0,
\end{equation}
that turns out to be satisfied by $f=f_k$ only for $\alpha=0$.  The
competition is apparently present also in the ordinary driven-damped
oscillator but in that case it is balanced by the $(2\pi f)^2$
contained in $V(f)$ at the numerator in such a way that the
nonmonotonic behavior of the energy can be explained only by the
resonance.  We finally note that another way to see competitive terms
in Eq. \eqref{SoluzSimp} is to rewrite it as
$\langle\dot{z_g}^2\rangle=\left(u_1/V+u_2/V \right)^{-1}$. In this
form we have the inverse of a sum of two terms that, for $f<f_k$,
depend in opposite way and with different powers on $f$. This clearly
gives raise to an extremal point also in the limit $f \ll f_k$, that is
consistent with the values of the fitted parameters.

We stress that these mechanisms substantially differ from the standard
resonance phenomenon, due to the presence of the damping term $f_s\sim
af^\alpha$. Indeed, an increase of $f$ induces the grains to adsorb
the injected energy in a faster vibrating motion, experiencing a
larger number of collisions (dissipative events) per unit of
time. This means that at higher $f$, both the energy input and the
energy output increase. The two phenomena compete and, because of
their different functional dependencies upon $f$, an extremal point
appears. We note that the maxima in the curves of
Fig.~\ref{fig:KeVariogN} and~\ref{fig:modVSdat} occur at frequencies
much smaller than the fitted $f_k$. Conversely, the ordinary
resonance always occurs at the fixed characteristic frequency $f_k$,
independently of the viscous coefficient. For $\alpha>0$,
the dissipation controls the position of the peak, reproducing the key
feature of molecular dynamic simulations.

\section{Comparison with previous studies}
\label{sec:comparison}

It is interesting to discuss some previous results
\cite{windows2015,windows2} about energy transfer optimization in
vibrated granular media to compare with those presented in this paper.
In \cite{windows2015} the authors present experimental and numerical
data for a vertically-shaken granular system in a 3D geometry. They
find a behavior of the kinetic energy $K$ versus the driving frequency
$f$ similar to the one found by us. Their study
is done for increasing $f$ and decreasing $A$ keeping constant the
shaker strength $S=(2\pi f A)^2/(gd)$. The main result is that there
is a specific combination of $f$ and $A$ that, at a given fixed $S$,
optimizes the energy transfer between the shaker and the granular
bed. This nonmonotonic trend is rationalized by noting the following
competitive effects: on the one hand, raising $f$ makes the maximum
rescaled acceleration $\Gamma=(2\pi f)^2A/g$ larger enhancing the
fluidization of the granular medium; on the other hand, decreasing
$A$ lowers the mean number of collisions with the shaker weakening the
interaction with the external source of energy.

Regarding \cite{windows2}, the authors study the same setup but
particularly concentrating on a 1D geometry, i.e. vertically-shaken
columns of single grains. In this situation, when $K$ is plotted
against $f$ at fixed $S$ they find many peaks at the integer multiples
of a specific frequency $f^*$. They interpret this phenomenon as a
resonant behavior: when the driving frequency is synchronized with the
typical time of detachment ($\tau^*=1/f^*$) of the granular column the
energy transfer from the shaker to the system is optimized.

In the following, we comment on some crucial evidences about the
differences between the results reported in our study and
those in~\cite{windows2015,windows2}.
 
\subsection{A different granular phase}

\begin{figure}[t!]
\includegraphics[width=0.35\textwidth]{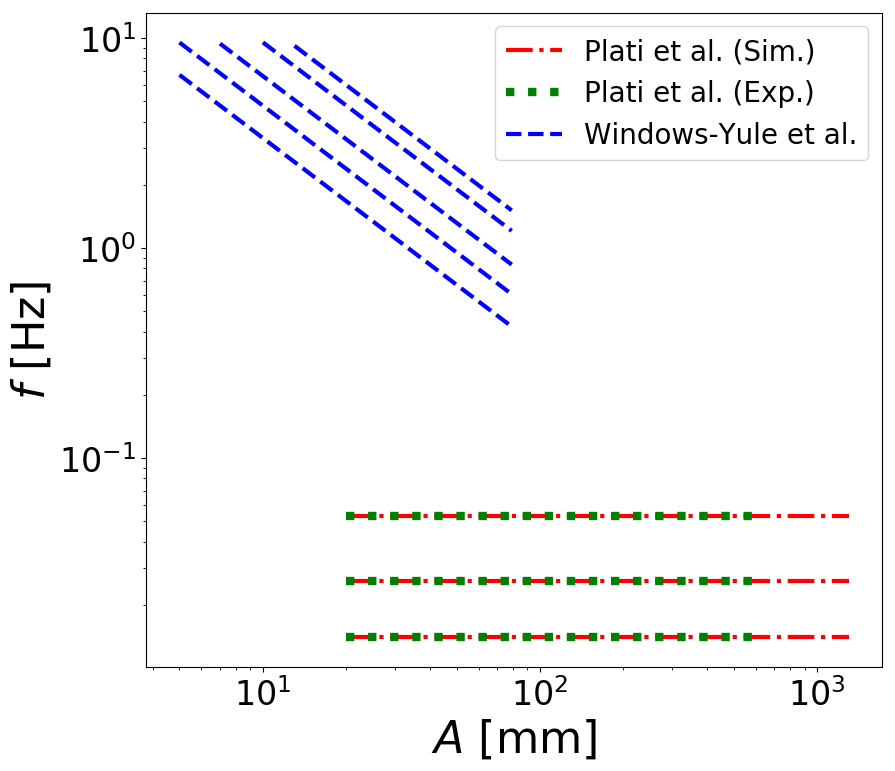}
\caption{Different regions of the parameter space investigated in our
  work with respect to Refs. \cite{windows2015,windows2}. } \label{sketch}
\end{figure}

Previous studies consider a
range of parameters which is completely different from the one
considered by us. More precisely, in our experiments/simulations we
consider $f \in[20,1300]$ Hz and $A=0.014, 0.026, 0.039, 0.053$ mm, so that $S \in[0.0003,1.15]$;
%and $\Gamma=(2\pi f)^2A/g \in [0.04,177.0]$;
on the contrary Refs. \cite{windows2015,windows2} investigate the
ranges $f\in[5,80]$ Hz and $A\in[0.6,8.0]$ mm, corresponding to $S \in
[0.9,11.5]$,
%and $\Gamma \in [1.0, 23.9]$,
see Figure \ref{sketch}.  They consider amplitudes $A$ much larger
than those studied in our system. For those values of $A$ the system
detaches from the driving plate reaching the usually called bouncing
bed state, where it is possible to define the time of free
flight $\tau^*$ responsible for the aforementioned resonant behavior.
We also stress that in \cite{windows2015} the granular medium is in a
dilute phase where every grain explores all the accessible space
whereas in \cite{windows2} the grains are arranged in a dense state
and each particle in the column remains in the same position with
respect to the others.

\begin{figure}[t!]
\includegraphics[width=0.35\textwidth]{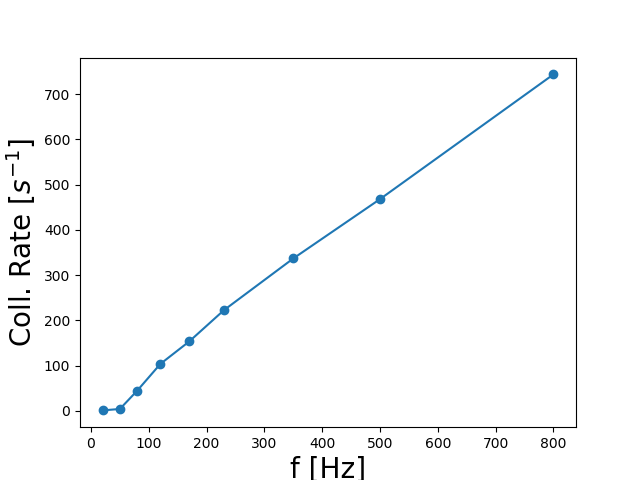}
\caption{Collision rate between grains and bottom plate as a function of $f$ with $A=0.026$ mm.} \label{collrate}
\end{figure}

In our article we show that a vibrofluidized state exists at much
smaller values of $A$ (see Fig. \ref{sketch}) which are sufficiently
small to keep the granular medium always in contact with the bottom of
the container. This is supported by the collision rate as a function
of $f$, plotted in Fig. \ref{collrate}. In our system, indeed, the
collision rate is an almost linearly increasing function of $f$, in
sharp contrast with the explanation of the phenomenon reported in
Refs. \cite{windows2015,windows2}. More precisely we have checked
(visually in the experiments and quantitatively in the simulations)
that, for all the driving parameters we explored, the grains are
always arranged in a dense packing where they vibrate around an almost
fixed position experiencing rare and slow rearrangements.  As a
consequence, in the considered range of parameters, the notion
of time of free flight is meaningless and the mechanism responsible
for the nonmonotonic behavior of $K$ has nothing to deal with a
resonant mechanism.  Indeed, at variance with the bouncing-bed state,
the permanent contact allows the oscillating driving plate to
continuously transfer kinetic energy to the granular medium as the
driving frequency is increased. In our case, the nonmonotonicity of
$K$ originates from a competition between driving energy and
dissipation, which both increase with frequency, but with different
powers of $f$.  In this sense, taking the point of view of the authors
of \cite{windows2015,windows2}, we observe a maximum of the internal
granular energy even if the input energy is increasing.

\subsection{A different role of dissipation}

\begin{figure}[t!]
\includegraphics[width=7cm]{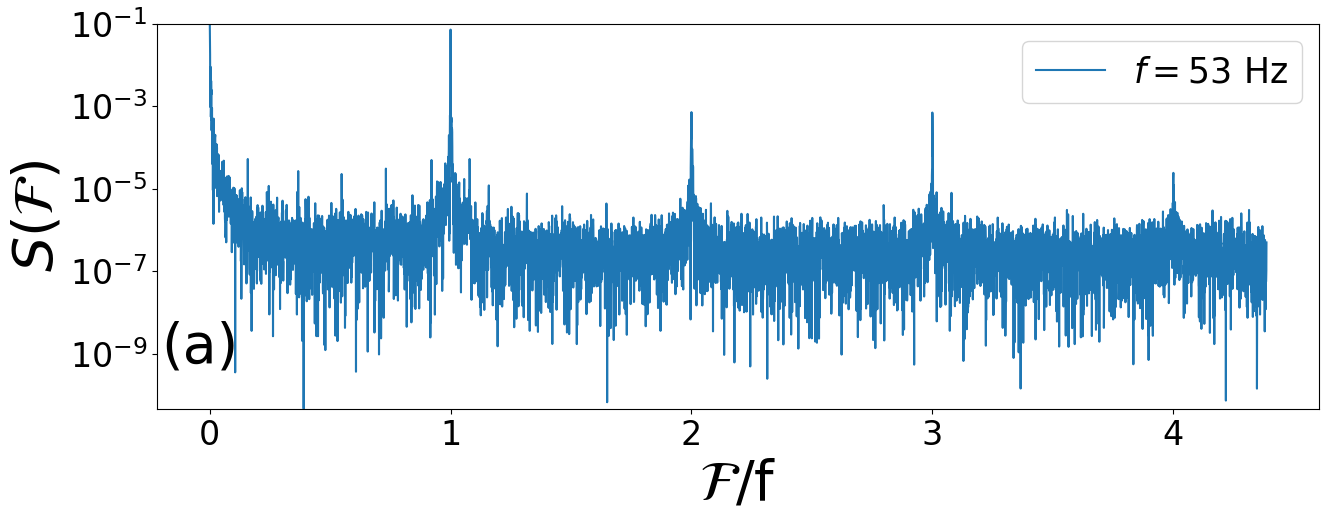}
\includegraphics[width=7cm]{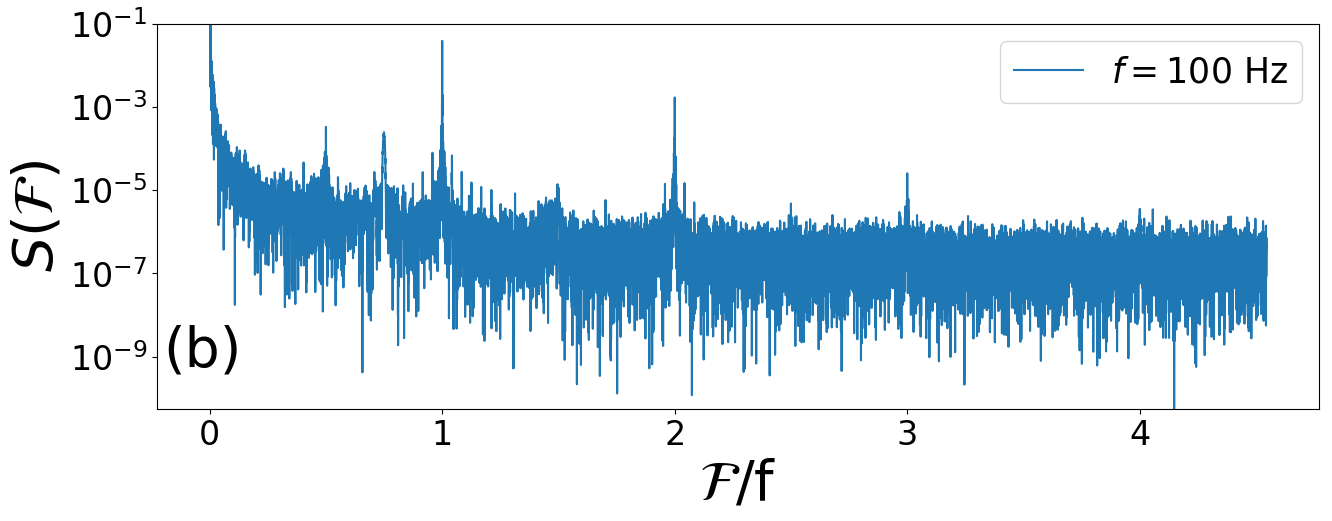}
\includegraphics[width=7cm]{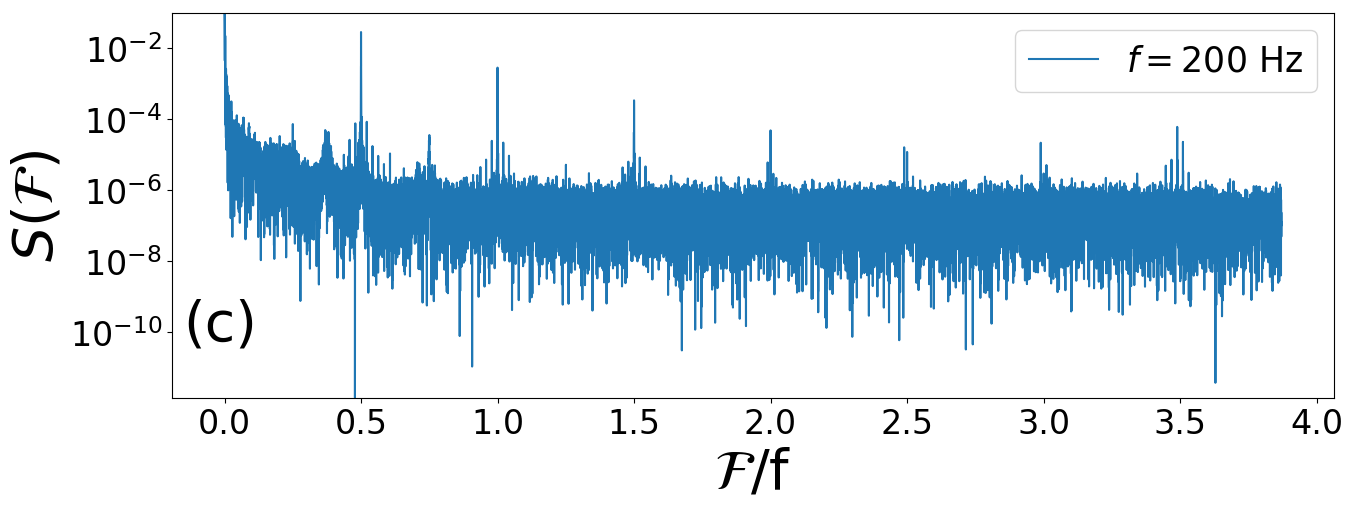}
\includegraphics[width=7cm]{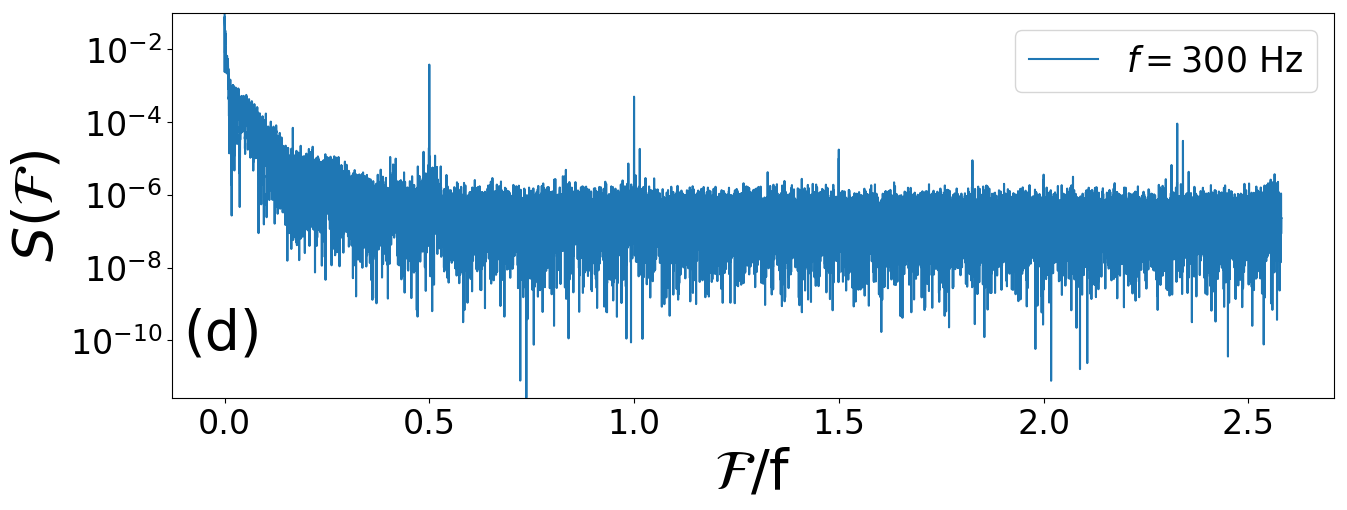}
\caption{Experimental data for the spectra of the vertical z-coordinate of the top plate for
  different shaking frequencies $f$ and $A=0.053$ mm.} \label{fig:spectra}
\end{figure}

In the bouncing bed state the optimal frequency $f^*$ is obtained 
when the vibration period $\tau^*$ is
synchronized with the flight time of the granular bed leading to Eq.(5) of Ref. \cite{windows2}
\begin{equation}
\tau^*= \frac{2\pi f A}{g}\left\{2+\beta\epsilon\exp\left[-\left(1-\epsilon^2 \right)\left(N-1\right) \right]\right\},
\label{eq:tof}
\end{equation}
where $g$ is the gravitational acceleration, $\epsilon$ the
restitution coefficient (i.e. the fraction of energy lost during a collision), $N$ the number of grains and $\beta$ a
fitting parameter that depends on the details of the experimental
setup.  Eq. \eqref{eq:tof} clearly shows that the resonant frequency
$f^*=1/\tau^*$ increases when the restitution coefficient is
decreased: the authors rationalize this observation with the fact that
more dissipated energy in collisions implies shorter flights,
i.e. smaller $\tau^*$ or higher resonant frequencies $f^*$.  In our
case, conversely, our peak frequency decreases when the
friction coefficient $\gamma_n$ of the normal force between the grains
(and between grains and borders) is increased. This is the evidence of
a very different mechanism at work: when we increase $\gamma_n$,
i.e. dissipation per single collision, the range of frequencies where
dissipation wins the competition is larger, and this corresponds to a
shift of the peak to the left.

\subsection{Spectra of the top plate}
\label{spectra}

In Fig. \ref{fig:spectra}, we show the spectra of the z-coordinate
(measured with a laser sensor, see \citep{gnoli2018} for details)
of the top plate for different values of the driving frequency $f$ and
$A=0.053$ mm. This figure sheds light on the relation between chaotic
motion and energy transfer in our system. We see that the most
coherent motion (pronounced peaks on the driving frequency and its
harmonics) is obtained for $f$ = 53 Hz (panel (a)), before the energy
maximum. At the driving frequency where our maximum in the energy
occurs (panel (b), $f$ = 100 Hz), we see that the peaks on the harmonics
are less pronounced and more broadband. For increasing driving $f$
(decreasing transferred energies), we see a more and more chaotic
spectrum with the appearance of distinct peaks at noninteger multiple
of $f$ (panel (c)) and the disappearance of the peaks in the harmonics
(panel (d)). Our scenario, therefore, is that of an increasing
chaoticity - with $f$ - of the granular dynamics, irrespective of the
energy transfer, i.e. both before and after the energy maximum. This
is quite different from the one observed in
\cite{windows2015,windows2} where the energy transfer to the system is
optimized at the frequency such that the granular system is more
regular and shows the most defined peaks in its spectrum.

\section{Conclusion}
\label{sec:conclusion}

Counter-intuitive phenomena, forbidden by
standard thermodynamic arguments, can occur in nonequilibrium systems.
Negative differential mobility and negative specific heat are typical
examples~\cite{benichou,sarracino,zia,brilliantov2}. This kind of
nonmonotonic behavior in the current-force relation is due to the
combination of competing mechanisms.  Here we have unveiled a similar
effect in a complex many-body system with dissipative interactions:
The interplay between external forcing and internal dissipation in
granular media can result in a nonmonotonic behavior of the system
kinetic energy as a function of the input energy, representing an
instance of negative specific heat. Moreover, our analysis explains
the important role played by the vibration frequency, triggering
specific dissipation mechanisms. These arise at different scales, from
rheological behavior to single-particle motion.  The observed
phenomenology may have a deep impact application in several fields
related to granular matter physics.

\acknowledgments{E. Lippiello, A. Puglisi and A. Sarracino acknowledge
  support from the MIUR PRIN 2017 project 201798CZLJ. L. de Arcangelis acknowledges
  support from project PRIN2017WZFTZP. L. de
  Arcangelis, E. Lippiello and A. Sarracino acknowledge support from
  VALERE project of the University of Campania
  ``L. Vanvitelli''. A. Plati, A. Puglisi and A. Gnoli acknowledge
  support of Regione Lazio through the Grant "Progetti Gruppi di
  Ricerca" N. 85-2017-15257.}

\appendix

\section{Simulation Details}
\label{app}

\begin{table}[t!]
\centering
\begin{tabular}{|c|c|} 
\hline 
$k_{n}^{a}$ & ($0.3-1.2)\times 10^{8}$ Pa \\ 
\hline
$k_{n}^{b}$ & 0.6 $\times 10^{8}$ Pa \\ 
\hline 
$k_{t}^{a}=k_{t}^{b}$ & 1.3$\times 10^{8}$ Pa \\
\hline
$\gamma_{n}^{a}=\gamma_n^{b}$ & 2.9$\times (10^{5}-10^{8})$ $ (\text{ms})^{-1}$ \\
\hline
$\gamma_{t}^{a}=\gamma_t^{b}$& 2.3$\times (10^{3}-10^{6})$ $ (\text{ms})^{-1}$ \\
\hline 
$\mu^{a}=\mu^{b}$ & 0.5\\ 
\hline
$dt$& 2.7$\times 10^{-6}$ s \\
\hline
\end{tabular} 
\caption{Numerical values for the material properties and the
  coefficients of the viscoelastic interaction. We use the index $a$
  for the grain-grain, grain-vane and grain-lid interactions while
  index $b$ is used for the grain-wall ones.}
\label{tab:TabParam}
\end{table}

Our molecular dynamics simulations are performed through LAMMPS
package \cite{LAMMPS} and the system geometry reproduces the
experimental setup of Ref.~\citep{gnoli2018}. The values of
$\langle \Omega\rangle$ and $K$ reported in the
main text are the result of a time average over $3\times 10^{7}$ time
steps (78 equivalent seconds) in the steady state.  Regarding the
interaction between the grains, we used the Hertz-Mindlin (HM) model
\cite{zhang,silbert,brilliantov} to solve the dynamics
during the contact. This viscoelastic model takes into account both the
elastic and the dissipative response to the mutual compression. In
addition, it includes in the dynamics not only the relative
translational motion but also the rotational one. The model is
described by the following equations:
\begin{equation}
  \vec{F}^{N}_{ij}= \sqrt{R_{ij}^{\text{eff}}}\sqrt{\xi_{ij}(t)}\left[(k_{n}\xi_{ij}(t)-m^{\text{eff}}\gamma_{n}\dot{\xi}_{ij}(t)) \vec{n}(t)\right], \nonumber
\end{equation}
\begin{equation}
  \vec{F}^{T}_{ij}=
\begin{cases}
  -\sqrt{R_{ij}^{\text{eff}}}\left[\vec{F}^{\text{hist}}_{ij}+m_{\text{eff}}\gamma_{t}\sqrt{\xi_{ij}(t)}\vec{g}^{T}_{ij}(t)\right] & \text{if} \; \; \; |\vec{F}^{\text{hist}}_{ij} | \le |\mu\vec{F}^{N}_{ij}|  \nonumber       \\ 
  -\dfrac{|\mu\vec{F}^{N}_{ij}|}{|\vec{g}^{T}_{ij}(t)|}\cdot \vec{g}^{T}_{ij}(t) & \text{otherwise}, \nonumber
  \end{cases}
\end{equation}
\begin{equation}
  \vec{F}^{\text{hist}}_{ij}=k_t\bigintssss_{\text{s(t)}}\sqrt{\xi_{ij}(t')}\vec{ds}(t').\nonumber
\end{equation}
These equations are written for two particles with radius $R_i$,
$R_j$, center positions $\vec{r}_i$, $\vec{r}_j$ translational
velocities $\vec{v}_i$, $\vec{v}_j$ and rotational velocities
$\vec{\omega}_i$, $\vec{\omega}_j$.  The relative velocity is defined
as $\vec{g}_{ij}= (\dot{\vec{r}}_{i}-\vec{\omega}_{i}\times
R_{i}\vec{n})-(\dot{\vec{r}}_{j}+\vec{\omega}_{j}\times R_{j}\vec{n})$
where
$\vec{n}=\left(\vec{r}_{i}-\vec{r}_{j}\right)/\left|\vec{r}_{i}-\vec{r}_{j}\right|$;
we call $\vec{g}_{ij}^{N}$ and $\vec{g}_{ij}^{T}$ the two projections,
respectively normal and tangential, to the surface of contact.  The
instantaneous normal compression is represented by
$\xi_{ij}(t)=R_{i}+R_{j}-|\vec{r}_i-\vec{r}_{j}|$ and its derivative
is $\dot{\xi}_{ij}(t)=|\vec{g}^{N}_{ij}|$.  During the contact, the
particles are subjected to a normal force $\vec{F}^{N}_{ij}$ and a
tangential one $\vec{F}^{T}_{ij}$; both these components have an
elastic and a dissipative contribution respectively characterized by
the coefficients $k_{n}$, $k_{t}$, $\gamma_{n}$, $\gamma_{t}$. In the
normal force $\vec{F}^{N}_{ij}$ we can see an elastic term that
derives from the hertzian theory of contact mechanics
\cite{popov} characterized by a nonlinear dependence on the
displacement.  The HM model is used to describe the interactions
between all elements of the simulation (the box, the vane and the
lid) considering the flat surfaces of the box as spheres with
infinite mass and radius.  Further discussions on the physical meaning
and estimation of the model parameters and on the choice of the
simulation time step $dt$ can be found in the Supplemental Material of
\cite{plati2019}. Here we report the numerical
values used for the present study (Tab. \ref{tab:TabParam}).


\begin{thebibliography}{99}
\bibitem{kv} Since we are mainly interested in the study of the total
  granular kinetic energy, we focus on the vane velocity fluctuations
  around the average, subtracting the contribution due to the external
  forcing imposed by the motor. As shown in the rest of the paper,
  $K_v$ is a good proxy for the global granular energy.
\bibitem{footnote1}
   We have verified that the presence of the vane does
not affect significantly the granular features.
\end{thebibliography}
\end{document}